\documentclass[english]{article}

\usepackage[colorlinks,bookmarksopen,bookmarksnumbered,allcolors=red]{hyperref}
\usepackage{xcolor}
\usepackage{graphicx}
\usepackage{geometry}
\usepackage{amsmath, amssymb, bm}
\usepackage{authblk}
\usepackage{placeins}
\usepackage{pdflscape}

\definecolor{bluegreen}{RGB}{46,141,131}
\definecolor{darkgreen}{RGB}{46,139,87}
\definecolor{darkred}{RGB}{219,7,61}
\definecolor{darkblue}{RGB}{0,0,137}

\title{\textbf{Dynamic prediction of an event using multiple longitudinal markers: a model averaging approach}}

\author[1]{Reza Hashemi}
\author[,2]{Taban Baghfalaki\thanks{Corresponding author: taban.baghfalaki@u-bordeaux.fr}}
\author[2]{Viviane Philipps}
\author[2]{Helene Jacqmin-Gadda}
\affil[1]{Department of Statistics, Razi University, Kermanshah, Iran}
\affil[2]{Univ. Bordeaux, INSERM, INRIA, BPH, U1219, F-33000 Bordeaux, France}

\date{}

\begin{document}

\maketitle

\begin{abstract}
Dynamic event  prediction, using joint modeling of survival time and longitudinal variables, is extremely useful in personalized medicine. However, the estimation of joint models including many longitudinal markers is still a computational challenge because of the high number of random effects and parameters to be estimated. In this paper, we propose a model averaging strategy to combine predictions from several joint models for the event, including one longitudinal marker only   or pairwise longitudinal markers. The prediction is computed as the weighted mean of the predictions from the one-marker or two-marker  models, with the time-dependent weights estimated by minimizing the time-dependent Brier score. This method enables us to combine a large number of predictions issued from joint models to achieve a reliable and accurate individual prediction. Advantages and limits of the proposed methods are highlighted in a simulation study by comparison with the predictions from well-specified and misspecified all-marker joint models as well as the one-marker  and two-marker  joint models. Using the PBC2 data set, the method is used to predict the risk of death in patients with primary biliary cirrhosis. 
The method is also used to analyze a French cohort study called the 3C data. In our study, seventeen longitudinal markers are considered to predict the risk of death.
\end{abstract}

\section{Introduction}
\label{s:intro}
Joint modeling of longitudinal variables and time-to-event is an attractive domain of research  \cite{henderson2000joint,guo2004separate,tsiatis2004joint}.
These models are especially useful for the prediction of the individual risk of a health event using repeated measures of longitudinal markers. { In this framework, they are a flexible alternative to the landmark approach  \cite{van2007,van2008dynamic}}.
An important advantage of this kind of { dynamic } prediction is that it can be updated dynamically whenever a new measure of a longitudinal marker is available. 
As a result, joint modeling has become an important tool in the current development of personalized medicine  \cite{andrinopoulou2021reflection}.

Joint models combine a mixed model for the change over time of the marker and a time-to-event model for the event risk, including functions of the marker trajectories as explanatory variables. 
From a computational point of view, several packages are available to fit such models either in the Bayesian (\texttt{JMBayes} \cite{rizopoulos2014r}) or the frequentist (\texttt{JM} \cite{rizopoulos2010jm}, \texttt{JoineR} \cite{philipson2012joiner}, \texttt{FrailtyPack} \cite{krol2017tutorial}) framework. Several packages allow to compute individual predictions. Extended joint models have been implemented to deal with competing events  \cite{elashoff2007approach, williamson2008joint}, or multiple and non-Gaussian longitudinal markers  \cite{dunson2005bayesian}. Papageorgiou et al. \cite{papageorgiou2019overview} presented an overview of these extended joint models.

Joint models have been applied to predict the onset or the progression of many diseases, such as cancers  \cite{desmee2017nonlinear}, cardiovascular diseases  \cite{tanner2021dynamic,andrinopoulou2015dynamic}, AIDS  \cite{barrett2017dynamic} and dementia  \cite{ben2022five,li2022joint}. Despite the vast number of potential longitudinal predictors and biomarkers for these diseases, most proposed prediction models rely on a small number of predictors. Indeed, the estimation of joint models with many longitudinal markers is highly challenging due to the large number of random effects and parameters to be estimated, which leads to very long computation times and some convergence issues. As each longitudinal marker can measure different aspects of the pathological process, naive selection of a few longitudinal variables can lead to models with poor predictive abilities.
To overcome these computational issues, some authors have proposed two-stage approaches where a multivariate mixed model is first estimated to obtain summaries of the longitudinal trajectories that are then introduced as time-fixed (the random effects)  \cite{Signorelli2021} or time-dependent (current expected value) \cite{mauff2020joint} explanatory variables in a time-to-event model. However, for high-dimensional or multi-modal markers, the estimation of the multivariate mixed model can also be untractable. Signorelli et al \cite{Signorelli2021} proposed more parsimonious latent process mixed models but did not correct for the bias at first stage due to informative dropouts following the event.
Gomon et al.  \cite{gomon2024dynamic} and Signorelli  \cite{signorelli2023pencal} propose an innovative landmark approach, implemented in the  pencal \cite{signorelli2023pencal} package, that utilizes penalized regression calibration to effectively manage numerous longitudinal covariates as predictors of survival. This method utilizes mixed-effects models to capture the trajectories of longitudinal covariates up to a predetermined landmark time. Subsequently, it employs a penalized Cox model
including the predicted random effects from the mixed models
to predict survival probabilities. Additionally, Devaux et al.  \cite{devaux2022random} present Dynforest, a novel technique for predicting individual risk. Dynforest expands the capabilities of competing-risk random survival forests to accommodate endogenous longitudinal predictors. In this approach, time-dependent predictors are transformed into time-fixed features at each node of the tree, facilitated by mixed models. These transformed features are then utilized as potential factors for dividing subjects into different subgroups. The Aalen-Johansen estimator is used to estimate the probability of individual events within each tree. 
Another approach applicable for continuous outcomes utilizes multivariate functional principal component analysis (FPCA) for dimension reduction and feature extraction from multiple longitudinal outcomes  \cite{li2019dynamic}. It then uses scores on these features as predictors in a Cox proportional hazards model to make predictions over time. Nevertheless, FPCA is not robust to informative dropouts.

In this work, we propose to compute predictions on the basis of several longitudinal markers by averaging predictions from several joint models, each including only one longitudinal marker or pairwise longitudinal markers. Model averaging has been proposed to increase predictive abilities and cope with model uncertainty, which is most often neglected after model selection  \cite{fletcher2018model}. Model averaging consists of estimating several possible models and combining their estimates or predictions using some weighting methods. In Bayesian model averaging, the quantities of interest (model parameters or prediction probabilities, for instance) are computed as the mean of the estimates from several candidate models weighted by the posterior probability of each model given the data  \cite{hoeting1999bayesian}. Various approximations, such as the Watanabe-Akaike information criterion WAIC  \cite{watanabe2013widely}, can be used to compute the posterior probability of each model. In the frequentist approach, the weights can be proportional to Akaike's information criterion (AIC)  \cite{buckland1997model} or achieved by minimization of the mean squared error  \cite{charkhi2016minimum}.

Andrinopoulou et al. \cite{andrinopoulou2017combined} applied model averaging to compute predictions from several joint models with different dependence structures between the event and the markers to avoid the selection of a particular dependence structure. In this work, the weights relied  on the relative likelihood of each model, which cannot be used to average predictions from joint models including different markers since the likelihoods are not comparable in this context. By combining machine learning and landmarking, Gonçalves et al. \cite{gonccalves2020model} created a super learner for dynamic predictions. The predicted probabilities were the weighted mean of the predicted probabilities of each learner, with the weights selected to minimize the mean squared error of the prediction. 

In this paper, we propose a model averaging approach to predict the probability of an event from repeated measures of several longitudinal markers by the weighted mean of the predictions issued from several joint models for the event and one or two longitudinal markers. This approach avoids estimating a joint model including all the markers that may be untractable as the number of markers increases. The time-dependent weights are computed to minimize the prediction error in each time window as measured by the dynamic Brier score \cite{blanche2015quantifying}. In this way, the weights account for the covariance between the markers. We call the proposed methods \texttt{MBSMA} as an abbreviation for "minimum Brier score model averaging".
{ We provide an R-package \texttt{DPMA}  (Dynamic Prediction by using Model Averaging) implementing the proposed methods. This package is available at \url{https://github.com/tbaghfalaki/MAJM}. }
The paper is organized as follows: Section 2 introduces individual predictions from joint models. Section 3 describes the proposed model averaging approach. In Section 4, the method is applied to predict the risk of death in the PBC data set { using seven biological longitudinal markers and the risk of death in the 3C cohort using seventeen longitudinal markers}. Section 5 presents a simulation study for comparing the behavior of the model averaging approach  with the predictions coming either from the one-marker joint models, the two-marker joint models, or from the all-marker joint model. The article ends with a discussion of the advantages and limitations of the proposed approach.

\section{Prediction from joint models }\label{sec2} 
Denote the time until the event occurs for subject $i$ by $T_i$ and the censoring time by $C_i$. As a result, the observed time will be $T^*_i=min(T_i, C_i)$, and $\delta_i $ will be the failure indicator $\delta_i=1_{T_i<C_i}$. Let $Y_{ijk}=Y_{ik}(t_{ijk})$ represent the longitudinal marker $k$ at time $t_{ijk}$ for the $i$th subject and the $j$th repeated measurement. The following observations are assumed to be equally distributed across a random sample of $N$ subjects: $\{T^*_i, \delta_i, Y_{ijk}, i=1, \cdots, N, j=1, \cdots, n_i, k=1, \cdots, K\}$.

\subsection{Joint model for longitudinal markers and time-to-event}
Joint models for the change over time of one longitudinal marker and the risk of one event combine a mixed model and a time-to-event model. In the shared random effect approach, the risk depends on the marker through the functions of the subject-specific random effects from the mixed model. In the latent class approach, the population is assumed to be split into several latent classes, and the time-to-event and mixed models are class-specific. After parameter estimation, both types of models can predict individual survival probabilities between times $s$ and $s+t$ given the information available. Our proposed model averaging approach can be applied to both types of joint models. In this paper, we focused on shared random effect models because this is the prevailing method with several packages available for one or several markers with various distributions in the exponential family.

By combining a proportional hazard model and a generalized linear mixed model, we considered a shared random effect joint model for a time-to-event and a longitudinal marker ${Y}_k$ that follows an exponential family distribution. The density of $Y_{ijk}$ has  the following form:
\begin{equation}
f_{Y_{ijk}}(y)=\exp{ \{ \psi_k^{-1} [\theta_{ijk} y - h_k(\theta_{ijk})] +C_k(y,\psi_k) \}}
\label{density}
\end{equation}
where $\psi_k$ is the dispersion parameter, and 
$h_k(.)$ and $C_k(.,.)$ are known functions. The formulation of the generalized linear mixed model is completed by
\begin{equation}
E(Y_{ijk}|\bm{b}_{ik})=  h_k'(\theta_{ijk}) = h_k'( \bm{W}_{ijk}^\top \bm{\beta}_k + \bm{Z}_{ijk}^\top \bm{b}_{ik}),
\label{mixed}
\end{equation}
where the link function $h_k'(.)$ in (\ref{mixed}) is the derivative of the function $h_k(.)$ in (\ref{density}); $\bm{W}_{ijk}$ and $\bm{Z}_{ijk}$ are vectors of covariates, possibly including time-dependent exogenous covariates, the time variable $t_{ijk}$ and interactions with $t_{ijk}$, such as $\bm{Z}_{ijk} \subset \bm{W}_{ijk}$. The random effects $\bm{b}_{ik} \sim N(0, \bm{B}_k)$. The proportional hazard model is defined by:
\begin{equation}
\lambda_i(t| \bm{b}_{ik})= \lambda_{0}(t) \exp{(\bm{X}_i^\top \bm{\gamma} +   g_k(\bm{b}_{ik},t)^\top \bm{\alpha})} \label{risk}
\end{equation}
where  $\bm{X}_i$ is a vector of time-fixed covariates and $g_k(\bm{b}_{ik},t)$ is a known function of the random effects $\bm{b}_{ik}$, the time and possibly fixed effects $\bm{\beta}_k$ that defines the dependence structure between the event and the marker. The likelihood for this model has the following form:
\begin{equation}
L(\bm{T}^{*},\bm{Y}_{k})=\prod_{i=1}^{N}\int L_{{Y}}(Y_{ik}|\bm{b}_{ik})\lambda({T}^{*}_{i}| \bm{b}_{ik})^{\delta_i} S({T}^{*}_{i}| \bm{b}_{ik}) f_{{b}}(\bm{b}_{ik})d\bm{b}_{ik},\qquad k=1,\cdots,K \label{likelihood}
\end{equation}  
where $L_{ \bm{Y}}(Y_{ik}|\bm{b}_{ik})$ is the product of the univariate conditional densities of $ Y_{ijk}$ given the random effects and $S({T}^{*}_{i}| \bm{b}_{ik}) $ is the conditional survival function.
This model can be estimated either by maximizing this likelihood with numerical integration on the random effects or using a Bayesian approach. It has been extended to multivariate longitudinal data by modeling several markers at the same time, allowing for correlation between marker-specific random effects:
$$\bm{b}_{i} \sim N(\bm{0}, \bm{B}),$$
where $\bm{b}_{i}$ is the stacked vector of $\bm{b}_{ik}$ and with the time-to-event sub-model specified by:
\begin{equation}
\lambda_{i}(t)= \lambda_{0}(t) \exp(\bm{X}_i^\top \bm{\gamma}+   \sum_{k=1}^K g_{k}(\bm{b}_{ik},t)^\top \alpha_{k}).\label{risk2}
\end{equation}
However, estimation of joint models with multiple markers is challenging because the number of parameters to be estimated and, more importantly, the number of random effects that define the size of the numerical integrals in (\ref{likelihood}) increase with the number of markers. In practice, this is often intractable with more than three or four markers and flexible structures for the random effects.

\subsection{Individual risk predictions}
Individual predictions of event risk are frequently provided by the joint models. For this purpose, the estimated model parameters are used to compute the probability of the event occurring in a time window $(s,s+t]$ given that the subject is free of the event at time $s$ and given the observations of the markers collected until time $s$:
\begin{eqnarray}\label{pi11}
\mathbb{\pi}_{i}(s,t) & = & P_{\bm{\xi}}(s \le T_{i}<s+t|T_{i}>s,\bm{\bm{\mathcal{Y}}}_{i}(s),\bm{\mathcal{X}}_{i}(s))
\nonumber \\
& = & 1-\int P_{\bm{\xi}}( T_{i}>s+t|T_{i}>s,\bm{\mathcal{Y}}_{i}(s),\bm{\mathcal{X}}_{i}(s),\bm{b}_{i}) \times f(\bm{b}_{i}|T_{i}>s,\bm{\mathcal{Y}}_{i}(s),\bm{\mathcal{X}}_{i}(s)) d\bm{b}_{i} \nonumber \\
& = & 1-\int P_{\bm{\xi}}( T_{i}>s+t|T_{i}>s,\bm{\mathcal{X}}_{i}(s),\bm{b}_{i}) \times f(\bm{b}_{i}|T_{i}>s,\bm{\mathcal{Y}}_{i}(s),\bm{\mathcal{X}}_{i}(s)) d\bm{b}_{i} \nonumber \\
& = & 1-\int \frac{S(s+t|\bm{b}_{i},\bm{\mathcal{X}}_{i}(s))}{S(s| \bm{b}_{i},\bm{\mathcal{X}}_{i}(s))} f(\bm{b}_{i}|T_{i}>s,\bm{\mathcal{Y}}_{i}(s),\bm{\mathcal{X}}_{i}(s)) d\bm{b}_{i},
\end{eqnarray}
where $P_{\bm{\xi}}$ is the probability distribution characterized by the vector of parameters $\bm{\xi}$ including all parameters in models specified by (\ref{density}), (\ref{mixed}) and (\ref{risk}) or  (\ref{risk2}) 
and depending on the vector of all covariates $\bm{\mathcal{X}}_{i}(s)$ and markers $\bm{\mathcal{Y}}_{i}(s)$ measured up to time $s$ for subject $i$,  that is,
$$
\bm{\mathcal{Y}}_{i}(s)=\{Y_{ik}(t_{ijk}); 0\le t_{ijk}\le s, j=1,\cdots,n_i, k=1,\cdots,K \},
$$
and
$$
\bm{\mathcal{X}}_{i}(s)=\{X_i, W_{ijk}; 0\le t_{ijk}\le s, j=1,\cdots,n_i, k=1,\cdots,K \}.
$$
These predictions are said to be "dynamic" because they may be updated with each new measurement of the markers. These prediction probabilities can be computed using the Bayesian method and MCMC  \cite{rizopoulos2012joint} or by Monte Carlo simulation and numerical integration  \cite{proust2014joint}.
In this work, joint models are estimated using the MCMC algorithm implemented in \texttt{JMBayes} \cite{rizopoulos2011dynamic}, and individual predictions are computed by the following Monte Carlo approach: parameters from the joint model are sampled from their posterior distribution given the trained data, $\bm{\xi}^{(m)}, m=1,\cdots,M$; then the random effects for subject $i$, $\bm{b}_{i}^{(m)}$, are sampled from $f(\bm{b}_{i}|T_{i}>s,\bm{\mathcal{Y}}_{i}(s),\bm{\mathcal{X}}_{i}(s),\bm{\xi}^{(m)})$ using a Metropolis-Hastings algorithm with independent proposals from a properly centered and scaled multivariate t distribution. Using $\bm{\xi}^{(m)}$ and $\bm{b}_{i}^{(m)}$, we can compute 
\begin{eqnarray}\label{gene}
    P_{\bm{\xi}}( T_{i}>s+t|T_{i}>s,\bm{\mathcal{X}}_{i}(s),\bm{\xi}^{(m)},\bm{b}_{i}^{(m)})=\frac{S(s+t|\bm{b}_{i}^{(m)},\bm{\mathcal{X}}_{i}(s),\bm{\xi}^{(m)})}{S(s| \bm{b}_{i}^{(m)},\bm{\mathcal{X}}_{i}(s),\bm{\xi}^{(m)})},~m=1,\cdots,M.
\end{eqnarray}
Finally, the mean of the $M$ sampled values is used to calculate the individual prediction probability:
$$1-\widehat{\mathbb{\pi}} _{i}(s,t) = \frac{1}{M}\sum_{m=1}^M \frac{S(s+t|\bm{b}_{i}^{(m)},\bm{\mathcal{X}}_{i}(s),\bm{\xi}^{(m)})}{S(s| \bm{b}_{i}^{(m)},\bm{\mathcal{X}}_{i}(s),\bm{\xi}^{(m)})}.$$

\subsection{Measures of prediction accuracy for time-dependent markers and outcome}
The evaluation of predictive abilities is an essential part of developing prediction models  \cite{steyerberg2009applications}. The leading indicators for evaluating the predictive abilities of prediction models are the Area Under the Receiver Operating Characteristics curve (AUC) and the Brier Score (BS). The area under the ROC curve measures the discrimination between future diseased and non-diseased subjects. The AUC quantifies the probability that the predicted probability of a randomly selected diseased subject is higher than the predicted probability of a disease-free subject. The accuracy of the predictions is measured by the BS, the mean squared error of the predictions.
 
These two indicators have been extended to deal with censored time-to-event and time-dependent markers. These time-dependent AUC and BS are functions of the prediction time $s$ and the horizon of prediction $t$. A nonparametric inverse probability of censoring weighted (IPCW) estimator has been proposed to account for right-censoring  \cite{blanche2015quantifying}. These indicators are computed for a set of selected times of prediction $s$ and clinically meaningful windows of prediction $t$.

The time-dependent AUC is  defined among subjects at risk at the prediction time $s$ as
$$AUC(s,t)=P(\mathbb{\pi}_{i}(s,t)>\mathbb{\pi}_j(s,t)|D_{i}(s,t)=1,D_{j}(s,t)=0,T_{i}>s,T_{j}>s),$$
where $D_{i}(s,t)=I_{\{s < T_{i} \le s+t \}}$ is the event indicator in the window $(s, s+t]$. Thus, this AUC is the probability that a randomly selected subject who had the event in the time window  $(s, s+t]$ has a higher predicted probability than a randomly selected subject free of event at $s+t$. Thus, the higher the better. The IPCW estimator accounting for right-censoring \cite{blanche2015quantifying} is:
\begin{eqnarray}\label{auc}
\widehat{AUC}(s,t)=\frac{\sum_i^N \sum_j^N 1_{\{\widehat{\mathbb{\pi}}_{i}(s,t)>\widehat{\mathbb{\pi}}_j(s,t)\}} D_{i}^{*}(s,t) (1-D_{j}^{*}(s,t)) \widehat{\Psi}_i(s,t) \widehat{\Psi}_j(s,t)  }{\sum_i^N \sum_j^N D_{i}^{*}(s,t) (1-D_{j}^{*}(s,t)) \widehat{\Psi}_i(s,t) \widehat{\Psi}_j(s,t) },
\end{eqnarray}

where $D_{i}^{*}(s,t)=I_{\{s < T_{i}^* \le s+t, \delta_i=1 \}}$ is the observed indicator of event in the window $(s, s+t]$ and the weights $\Psi_i(s,t)$ are estimated by

$$
\widehat{\Psi}_i(s,t) = \frac{1_{ \{T_i^*>s+t \}}}{\widehat{G}(s+t|s)} + \frac{1_{\{s<T_i^* \le s+t  \} }\delta_i}{\widehat{G}(T_i^*|s)},
$$
where $ \widehat{G}(u|s)$ is the conditional probability of being not censored at $u$ conditionally on being not censored at $s$,  $ \widehat{G}(u|s)=\widehat{G}(u)/\widehat{G}(s)$  and $\widehat{G}(u) $ is the Kaplan-Meier estimator of the probability to be uncensored at $u$. Thus the weight is the inverse of the conditional probability to be uncensored at  time of the event for incident cases observed in the window of prediction and at $s+t$ for subjects still at risk at the end of the window.

The AUC is a clinically meaningful indicator, but it measures only the discrimination between diseased and non-diseased subjects. Thus, the evaluation of the predictive abilities of a model is usefully completed by the computation of the BS, which measures the quadratic error of prediction. It is defined among subjects at risk at time $s$ as   \cite{blanche2015quantifying},  
$$BS(s,t)=E[\left(D_i(s,t)-\mathbb{\pi}_i (s,t)\right)^{2}|T_i>s],$$
and its IPCW estimator accounting for right-censoring is defined by:
\begin{eqnarray}\label{BS}
BS(s,t)&=& \frac{1}{N(s)} \sum_{i=1}^N  \widehat{\Psi}_i(s,t) (D_{i}^{*}(s,t)-\widehat{\mathbb{\pi}}_i (s,t))^{2}\\\nonumber
&=&\frac{1}{N(s)}  \{\sum_{\{i|T^*_i>s+t\}}\frac{\widehat{\mathbb{\pi}}_i(s,t)^2}{\widehat{G}(s+t|s)} 
+\sum_{\{i|s<T^*_i<s+t, \delta_i=1 \}}\frac{(1-\widehat{\mathbb{\pi}}_i(s,t))^2}{\widehat{G}(T^*_i|s)} \},\\\nonumber
\end{eqnarray}
where $N(s)$ is the number of subjects at risk at $s$ (subjects with $T_i^*>s$).

\section{Model Averaging}
We propose computing the individual probability of an event between times $s$ and $s+t$ as the weighted mean of the $K$ predicted probabilities provided by $K$ joint models for the event risk and one of the $K$ markers:
\begin{equation}\label{pi}
\widehat{\mathbb{\pi} } _{i}(s,t) = \sum_{k=1}^K\widehat{\mathbb{\pi} } _{ik}(s,t) w_k(s,t),
\end{equation}
where $\widehat{\mathbb{\pi} } _{ik}(s,t)$ is  the dynamic prediction   based on the joint model for marker $k$. We call this model averaging the one-marker MA.
Given that the data used for each joint model are partly different (the markers are different), the weights can not be defined as an approximation of the model probability given the data as in 
Andrinopoulou et al. \cite{andrinopoulou2017combined}.

To better account for marker dependence, we propose an alternative model averaging approach that consists in combining prediction of joint models with two-marker (in the spirit of Fieuws and Verbeke's pairwise approach \cite{fieuws2004joint}). For this purpose, $K$ in equation \eqref{pi} should be replaced by $\binom K2$ and 
 $\widehat{\mathbb{\pi} } _{ik}(s,t)$ is the dynamic prediction based on the 
 $k$th joint models including two markers. We call this model averaging the two-marker MA.
 
 We propose to select the weights which minimize the prediction error over the considered window of prediction $(s,s+t]$ of the overall predicted probabilities defined by (\ref{pi}). 
The IPCW estimator of the Brier score,  $BS(s,t)$,  accounting for right censoring  \cite{blanche2015quantifying} is used to measure prediction error. Replacing $\mathbb{\pi}_i(s,t)$ by (\ref{pi}) 
in the IPCW estimator (\ref{BS}), 
we obtain:
\begin{eqnarray}\label{BSw}
BS_w(s,t)
&=& \frac{1}{N(s)}  \{\sum_{\{i|T^*_i>s+t\}}\frac{(\sum_{k=1}^{\mathcal{K}} \widehat{\mathbb{\pi} } _{ik}(s,t) w_k(s,t))^2}{\widehat{G}(s+t|s)} \\
&+&\sum_{\{i|s<T^*_i<s+t, \delta_i=1 \}}\frac{(1- \sum_{k=1}^{\mathcal{K}} \widehat{\mathbb{\pi} } _{ik}(s,t) w_k(s,t))^2}{\widehat{G}(T^*_i|s)} \}, \nonumber
\end{eqnarray}
where ${\mathcal{K}}=K$ or $\binom K2.$
The model- and time-specific weights are the value of $w_k(s,t)$ that minimizes (\ref{BSw}) 
over the learning data set with the constraints of non-negativity and that their sum is equal to one. Thus
$$\widehat{w}_k(s,t)=arg min_{w_k(s,t)\in\mathcal{W}} BS_w(s,t),$$ 
where $\mathcal{W}=\{w_k(s,t)\in [0,1], ~k=1,\cdots,{\mathcal{K}},~ \sum_{k=1}^{\mathcal{K}} w_k(s,t)=1\}$.
Practically this minimization can be performed by using, for example, the \texttt{Rsolnp} package in $\it R$, which is based on the general non-linear optimization using the augmented Lagrange multiplier method by solving quadratic sub-problems \cite{ye1988interior}.\\
The method proposed  by Buckland et al. \cite{buckland1997model}, for  model averaging, is used to estimate the standard error of $\widehat{\mathbb{\pi} } _{i}(s,t)$ of equation \eqref{pi} as follows:
\begin{eqnarray}\label{var1}
\hat{\varsigma}_i(s,t)=\sum_{k=1}^{\mathcal{K}} \widehat{w}_k(s,t) \{\widehat{\mathcal{Z}}_{ik}(s,t)^2+\widehat{\mathcal{V}}_{ik}(s,t)\}^{1/2},
\end{eqnarray}
where $\widehat{\mathcal{Z}}_{ik}(s,t)$ and $\widehat{\mathcal{V}}_{ik}(s,t)$ are estimates of the $k$th model's bias and variance for landmark time $s$ and prediction window $t$, respectively. Also, $\widehat{\mathcal{V}}_{ik}(s,t)$ can be obtained by fitting model $k$ and using the standard approaches for estimating the variance of unknown parameters, and $\widehat{\mathcal{Z}}_{ik}(s,t)$  can be obtained as follows \cite{fletcher2018model}:
$$\widehat{\mathcal{Z}}_{ik}(s,t)=\widehat{{\pi} } _{ik}(s,t)-\widehat{\mathbb{\pi} } _{i}(s,t).$$
 A natural alternative to \eqref{var1} is proposed as follows \cite{burnham1998practical}:
\begin{eqnarray}\label{var2}
\hat{\hat{\varsigma}}_i(s,t)=\biggl\{\sum_{k=1}^{\mathcal{K}} \widehat{w}_k(s,t) \{\widehat{\mathcal{Z}}_{ik}(s,t)^2+\widehat{\mathcal{V}}_{ik}(s,t)\}\biggl\}^{1/2}.
\end{eqnarray}

\section{Applications}\label{section4}
\subsection{Application 1: PBC2 data}
In this section, we used the proposed model averaging approach to predict the risk of death in primary biliary cirrhosis patients from the PBC2 data set \cite{lin2002modeling} using repeated  measures of seven biological markers: spiders (binary), albumin (in mg/dl), log(alkaline) (alkaline phosphatase in U/liter), log(SGOT) (in U/ml), log(platelets) (platelets per cubic ml/1000), log(prothrombin) (prothrombin time in seconds) and log(serum bilirubin) (serum bilirubin in mg/dl). The sample consists of 312 patients enrolled in clinical trials at the Mayo Clinic from 1974 to 1984. Patients were followed up for 8.19 years in the median (maximum 14.3), with a median of 5 visits (maximum 16). In this analysis, the event of interest is death without transplantation, while subjects alive at the end of the study or transplanted are considered right-censored ($44.87\%$ dead). The complete data can be found in many R-packages, for example \texttt{joineRML}, with description in \url{https://rdrr.io/cran/joineRML/man/pbc2.html}.

We estimated several joint models for the risk of death, each including one,  two  or all the longitudinal markers. In all models, the sub-model for the time to death depends on the current value of the included markers and is adjusted for the treatment group (drug) and the patient's age at enrollment. The change over time of the markers is described using linear mixed models for the quantitative markers and a logistic mixed model for spiders, each including a linear time-trend with a random intercept and a random slope. 
A B-splines approach is  used to model the baseline hazard function $\lambda_0(.)$ of equation  \eqref{risk2} \cite{rizopoulos2012joint}.

The predictive abilities of each model, the one-marker MA and the two-marker MA for the 2-year risk of death using repeated measures of the markers collected until the landmark times 0, 2, 4, 6, 8, and 10 years are assessed through the AUC and BS. The estimated parameters from the time-to-event sub-model are presented in Table A. 1 in the web supplementary material A. This table shows the significant association of the event time with Spiders, Albumin, log(Prothrombin) and log(SerBilir) such that the first two have negative associations and the others have positive associations. The estimates for the regression coefficients and the variance of the errors of the longitudinal sub-model, and the covariance matrix of the random effects can be found in Tables A. 2 and A. 3, respectively.
In implementing the Bayesian approach for the model including the seven markers using the R package \texttt{JMbayes}, a large number of iterations of MCMC were required to attain convergence. Thus, two parallel MCMC chains were run for 200000 iterations each.
Then, we  discarded the first 100000 iterations as pre-convergence burn-in and retained 100000 for the posterior inference; thin was considered to be equal to 100. For checking convergence of the MCMC chains, we have used the Gelman-Rubin-Brooks diagnostic test  \cite{brooks1998general}. By contrast, the convergence of the one-marker and two-marker joint models, were obtained with the default number of iterations and burn-in of the package ($n.iter=28000,
n.burnin=3000$).
 
\begin{figure}[ht]
\centering
\includegraphics[width=20cm]{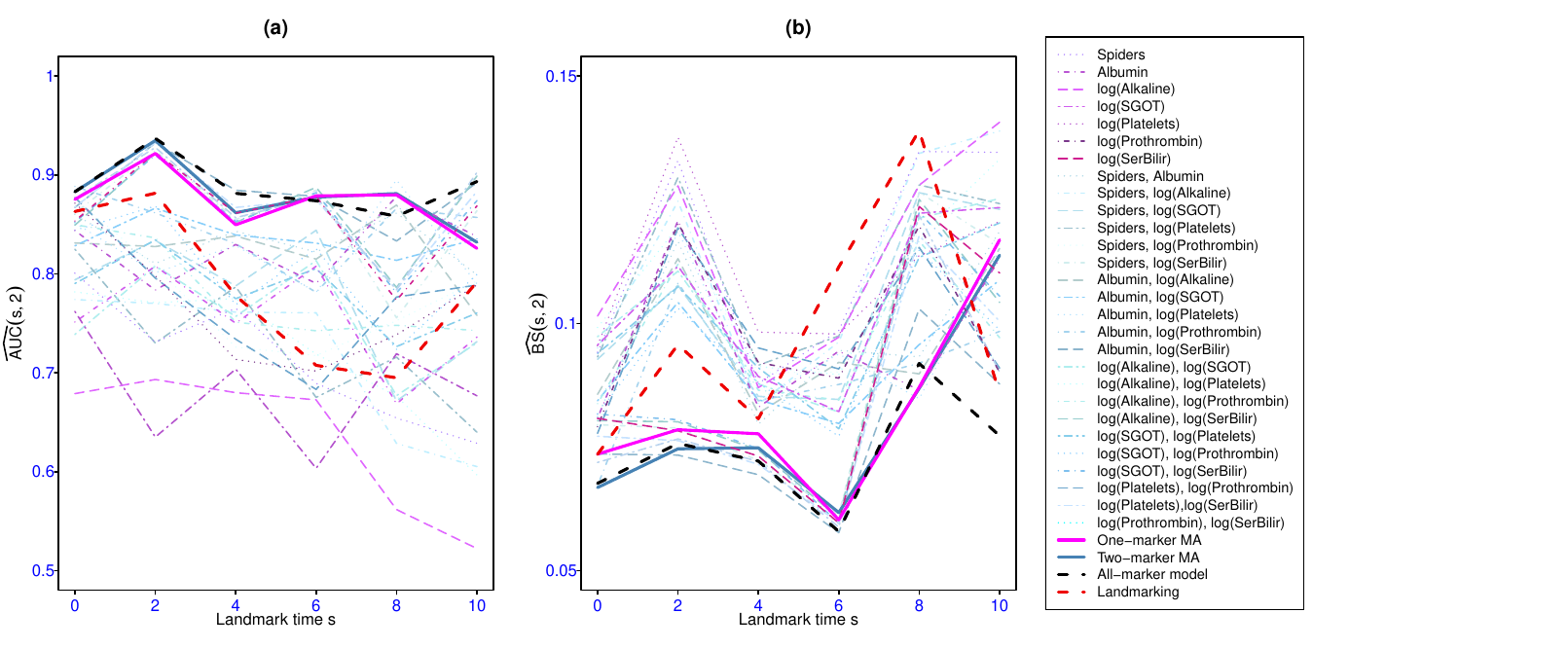}
\vspace*{-.1cm} \caption{\label{fig1}  
 AUC (panel a) and Brier score (panel b) computed by 5-fold cross-validation for landmark times $s=0,2,4,6,8,10$ years and prediction windows of 2 years for the comparison between the one-marker joint models,  the two-marker joint models, the all-marker joint model,  the one-marker MA and the two-marker MA   for PBC2 data.
 }
\end{figure}

\begin{landscape}
\begin{figure}[ht]
\centering
\includegraphics[width=22cm]{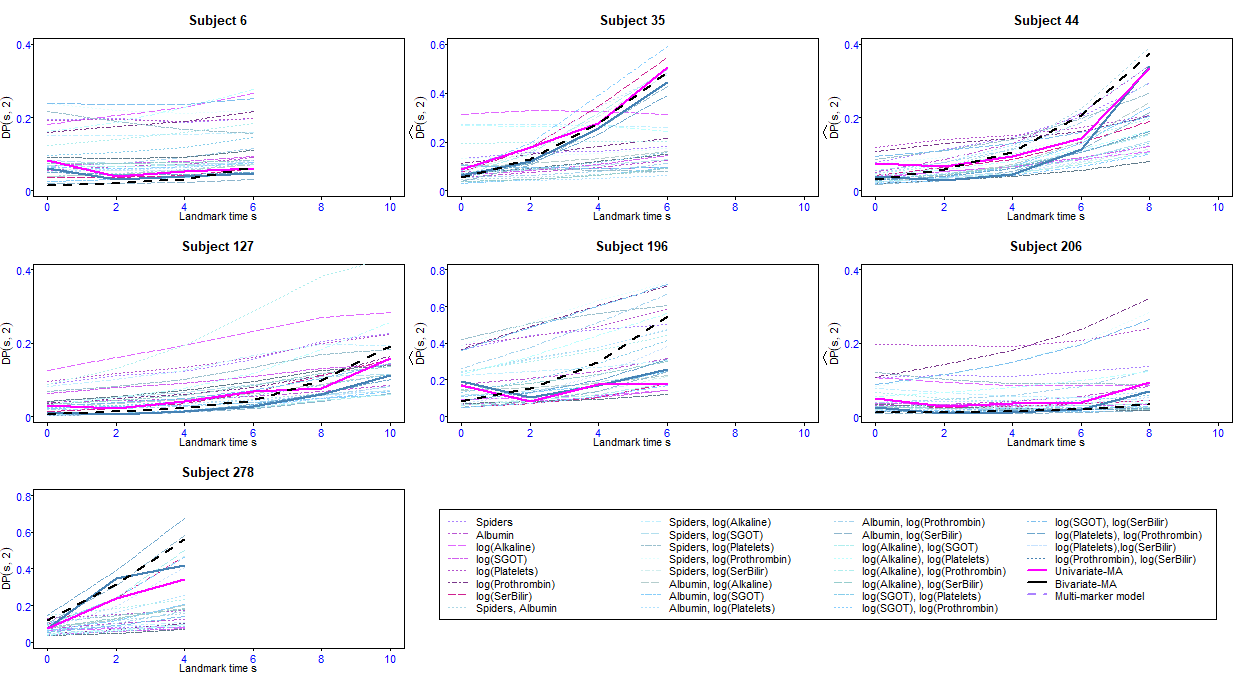}
\vspace*{-.1cm} \caption{\label{ind}  
Dynamic prediction computed by 5-fold cross-validation for landmark times $s=0, 2, 4, 6, 8, 10$ years and prediction windows of 2 years for the comparison between the one-marker joint models,  the two-marker joint models, the all-marker joint model,  the one-marker MA and the two-marker MA for some randomly selected individuals of PBC2 data. Subjects $6, 35, 44$ and $196$ died at $6.853, 7.794, 9.385$ and $6.995$, respectively; and subjects  $127, 206$ and $278$    were censored 
$10.743,  8.006$ and $5.465$, respectively.}
\end{figure}
\end{landscape}

\begin{figure}[ht]
\centering
\includegraphics[width=16cm]{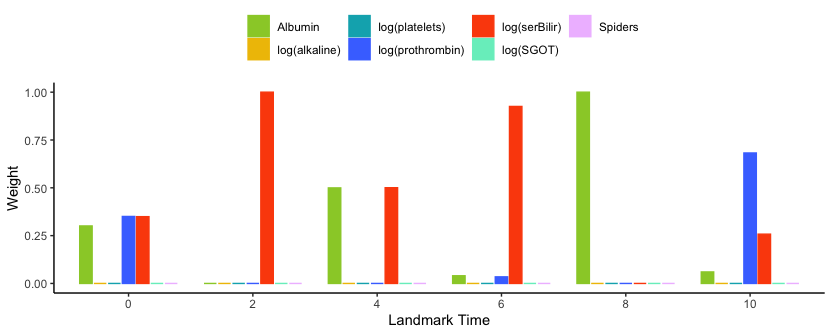}
\vspace*{-.1cm} \caption{\label{figweight}  
The distribution of weights for the landmark times $s=0, 2, 4, 6, 8, 10$ years and prediction windows of 2 years for the one-marker MA using PBC2 data.
 }
\end{figure}

Figure \ref{fig1} and Table A. 4 compare the 2-year predictive abilities measured by the AUC and the BS according to the landmark times for the all-marker joint model, the one-marker MA and the two-marker MA, the seven one-marker joint models, $\binom 7 2=21$ two-marker joint models { and the landmarking with last observation carried forward (LOCF) using the R package \texttt{Landmarking}  \cite{landmarking}}. The AUC and the BS were computed by 5-fold cross-validation using $80\%$ of the sample as the learning set and the remaining $20\%$ as the validation set. The weights for the model averaging estimator were  estimated on the learning set. 
According to the mean value of the AUC over the landmark times, the five best individual predictions were ordered as the following: all-marker joint model (JM), joint modeling of Albumin and log(SerBilir), two-marker MA, one-marker MA, and joint modeling of log(Prothrombin) and log (SerBilir). According to the mean BS, the order is unchanged  except that the bivariate joint model including log (prothrombin) and log (serbilir) appears slightly better than one-marker MA.
The values of  AUC and BS for these models  were very close at all landmark times (sometimes indistinguishable) and most often better than the best one-marker model and the other two-marker joint models. { Additionally, the predictive abilities of the landmarking approach with LOCF are poorer than the all-marker joint model and the approaches based on the model averaging.}
Figure \ref{ind}  shows the individual dynamic predictions for the all-marker joint model, the one-marker joint models, the two-marker joint models, the one-marker MA and the two-marker MA for some randomly selected individuals of PBC2 data. Although the dynamic prediction based on most of  the one-marker joint models and two-marker joint models were quite different, for some individuals, the estimated dynamic predictions based on 
the all-marker joint model,
 two-marker MA and one-marker MA  were close to each other except for subjects 196 and  278 at the last landmark times.
Table A. 5 displays the values of the individual dynamic predictions as well as their standard deviations.\\ 
{ Figure \ref{figweight} illustrates the weight distribution for various landmark times and a 2-year prediction window for the one-marker MA. The first significant point in this figure is the varying behavior of weights at different landmark times. Also, by comparing the estimates of Table A. 1, we can conclude that only significant markers may have a non-zero weight. Also, Figure A. 1 shows a similar plot for the two-marker MA. This figure also confirms that the significant markers have greater weights.  }

\subsection{Application 2: 3C study}
{In this section, we analyzed a subset of the Three-City (3C) Study  \cite{antoniak2003vascular} which is a French cohort study in which participants aged 65 or older were recruited from three cities (Bordeaux, Dijon, and Montpellier) and followed for over 10 years. The data includes socio-demographic information, general health information, and scores from several cognitive tests. The application only includes the center of Dijon and Bordeaux were MRI exams were conducted at years 0, 4 and also  years 10 for the center of Bordeaux. Except for the MRI markers, the other longitudinal markers were measured at baseline and at each follow-up visit (years 2, 4, 7, 10, 12, and also 14 and 17 in Bordeaux only).
In this study, we aimed at developing a dynamic prediction model for the risk of death. Exact time of death has been collected all along the follow-up.\\
We considered a subsample of $N = 1968$ subjects who had at least one measurement for each of the longitudinal markers. A total of 482 subjects died during the follow-up, and Figure A. 2 displays the Kaplan-Meier survival function during the follow-up period.
We considered seventeen longitudinal markers as potential predictors of death: three cardio-metabolic markers (body mass index (BMI), diastolic blood pressure (DBP), and systolic blood pressure (SBP)), the total number of medications (TOTMED), depressive symptomatology measured using the Center for Epidemiologic Studies-Depression scale (CESDT, the lower the less depressed), functional dependency assessed using Instrumental Activity of Daily Living scale (IADL, the lower the less dependent), four cognitive tests (the Benton visual retention test
 (BENTON, number of correct responses among 15), the Isaac set test of verbal fluency (ISA, total number of words given in 4 semantic categories in 15 seconds), the trail making tests A and B (TMTA and TMTB, number of correct moves by minute); the total intracranial volume (TIV)) and four biomarkers of neurodegeneration, including white matter volume (WMV), gray matter volume (GMW), left hippocampal volume (LHIPP) and right hippocampal volume (RHIPP); two markers of vascular brain lesions including volumes of White Matter Hyperintensities in the periventricular (Peri) and deep (Deep) white matter.
Figure A. 3 shows individual trajectories of the longitudinal markers in the 3C study, illustrating time-dependent variables. The data were pre-transformed using splines to meet the normality assumption of linear mixed models
 \cite{devaux2022random,proust2015estimation}.
The minimum, maximum, and median number of repeated measurements for SBP, DBP, CESDT, BENTON, ISA, TOTMED, and IADL are 1, 8, and 6, respectively. For BMI and TMTA, the minimum, maximum, and median number of repeated measurements are 1, 7, and 5, respectively. The minimum and maximum number of repeated measurements for TMTB is the same as those for TMTA, but the median is 4. The minimum, maximum, and median number of repeated measurements for WMV, GMW, TIV, RHIPP, and LHIPP are 1, 3, and 2, respectively. For Peri and Deep (which were not measured at year 10), there are at most two repeated measures, with a median of 2.
The analyses were also adjusted for age at baseline, education level, sex and diabetes status at baseline.
\\
Table A. 6 shows the estimated association parameters from the one-marker joint models on 3C data. The results show that 
GMV, HIPPR, HIPPL, CESDT, BENTON, ISA, TOTMED, TMTA, TMTB, and IADL are significantly associated with the death risk. As discussed in Section 4.1, Figure A. 4 shows that only the significant markers may have a non-zero weight for the MA.\\
{We compared the predictive abilities of different methods for predicting death at landmark times $s=0, 5, 10$ years,
with a prediction window of $5$ years. These methods included one-marker MA, the two-stage approach combining multivariate FPCA and Cox model as implemented in MFPCCox, the random survival forest for longitudinal predictors implemented in Dynforest, the penalized landmark approach with regression calibration implemented in pencal and landmarking with LOCF. For Dynforest, as suggested by the authors, we optimised the number of predictors considered at each split (mtry) and used the default values for the other tuning parameters.\\
The AUC and Brier score computed on the validation set presented in Table \ref{auc3c} demonstrate that one-marker MA had the best predictive abilities for the three
landmark times, quite close of MFPCCox. The landmark approach with LOCF had lower performances because it does not deal with measurement error. Pencal does not provide predictions at time zero because the mixed models can not be estimated with a single measurement. Also, as Pencal requires an identical structure for all mixed models, we used a linear time trend for all markers; this led to good result at time $5$ but poor at time $10$. Although Dynforest is able to provide a prediction at time $0$, AUC and BS were very bad because the prediction of $2$ or $3$ individual random effects using a single measure is probably very poor. Moreover, the lower performances of Dynforest may be due to the use of the logrank test as splitting rules that assumes proportional hazard over the time range while the two landmark approaches and the one-marker MA allows the parameters or the weights to depend on the landmark time. We conducted all the analyses presented in Table \ref{auc3c} using a MacBook Pro 2020 outfitted with Apple’s cutting-edge M1 chip and boasting 256GB of storage capacity. The computational time for the MA approach, without parallelization, amounts to 224 minutes. However,  this duration could be significantly reduced through parallelization. 
In comparison, the computational times for MFPCCox, landmarking, and pencal are 0.195, 0.026, and 0.523 minutes, respectively. Meanwhile, the computational time for DynForest, considering mtry in the range of 1 to 17 and optimizing its values for computing the risk prediction, is 38.030 hours.
}

\section{Simulation study}
A simulation study was performed to compare the performance of the proposed model averaging methods with the predictions provided by the all-marker joint model, the one-marker joint models  and the two-marker joint models. For this purpose, different scenarios were considered. In this section, the design of the simulation studies was described first, followed by various scenarios and the results.

\subsection{Design of simulation}
Data were generated using all-marker joint models with $K=2,3$ or $K=7$ longitudinal markers and one event, according to four different scenarios. In all scenarios, the longitudinal markers were assumed to be Gaussian or binary and generated according to a multivariate mixed effect model. The time-to-event was simulated according to a proportional hazard model with constant baseline hazard and depending on the current value of the $K$ markers. More specifically, the data generation models for the first three scenarios for the Gaussian markers are the following:
\begin{eqnarray}\label{gau}
Y_{ik}(t)&=&m_{ik}(t)+\varepsilon_{ikt}\\
&=&\beta_{0k}+\beta_{1k}t+b_{0ki}+b_{1ki} t+\varepsilon_{ikt},\nonumber
\end{eqnarray}
with $\varepsilon_{ikt} \sim N(0, \sigma_k^2)$ 
and for the binary markers is the following:
\begin{eqnarray}\label{bin}
\rm{logit} \left( P(Y_{ik}(t)=1|  b_{0ki},b_{1ki}  ) \right)&=&
m_{ik}(t)\\ &=&\beta_{0k}+\beta_{1k}t+b_{0ki}+b_{1ki}t, \nonumber
\end{eqnarray}
where $\bm{b}_i=(b_{01i},b_{11i},...,b_{0Ki},b_{1Ki})^{\top} \sim N(\bm{0},\bm{B})$, $i=1,\cdots,N,~k=1,\cdots,K$. The time-to-event sub-model is specified by
$$\lambda(t)=\lambda_0 \exp (\sum_{k=1}^K \alpha_k m_{ik}(t)).$$
For each scenario, 100 data sets of $N=1000$ subjects were generated and
split into a learning sample of $80\%$ and a validation sample of $20\%$. 
Marker measures were generated at times $t=0,0.2,0.4,\cdots,2$ and the time-to-event was administratively censored at $t=2$ in addition to a Weibull censoring. 
 We estimated the K joint models for the event and one longitudinal marker (one-marker models), $\binom K2$  joint models for the event and two  longitudinal markers and the well-specified K-marker joint model using the R-package \texttt{JMbayes} on each learning sample. The package provided the individual probabilities of the event
 for the landamark times $s=0,0.5,1,1.5$  and a prediction window $t=0.5$ for each subject in the learning set and each model.
The weights of the proposed model averaging method were then computed by minimizing the Brier score on the learning sample for each prediction window. Finally, the individual predictions were calculated for subjects from the validation data set using estimates from each model and the model averaging methods for the same prediction windows.  The one-marker joint models,  the two-marker joint models, the all-marker joint model,  the one-marker MA and the two-marker MA  were compared according to the AUC and the MSE  or the Brier score on the validation sample.
The AUC and BS criteria were given by \eqref{auc} and \eqref{BS}, respectively, and the true value of $\mathbb{\pi}_i (s,t)$ was computed using the true values of the parameters. 
The mean square error is given by $MSE(s,t)=1/N\sum_{i=1}^{N} (\widehat{\mathbb{\pi} } _{i}(s,t)-\mathbb{\pi}_i (s,t))^2,$ where $\widehat{\mathbb{\pi} } _{i}(s,t)$ is an estimate of \eqref{pi11}.

\subsubsection{Scenario 1: Three Gaussian markers }
In the first scenario, we considered $K=3$ Gaussian markers generated from equation \eqref{gau} such that $\bm{\beta}_k=(\beta_{0k},\beta_{1k})=(0,-1)$ and $\sigma_k^2=0.5,~k=1,2,3$. In this scenario, the following structures for $\bm{B}$ and values for $\bm{\alpha}=(\alpha_1,\alpha_2,\alpha_3)$ were investigated: 
\begin{description}
  \item[Scenario I with independent markers:] At first, a block-diagonal covariance matrix of the random effects:
  $$\bm{B}=\begin{pmatrix}
\bm{B}^* & \bm{0} & \bm{0}\\
  & \bm{B}^* & \bm{0}\\
 &  & \bm{B}^*
\end{pmatrix},$$
where
\begin{eqnarray}\label{c}
\bm{B}^*=\begin{pmatrix}
1 & 0.5\\
0.5  & 1  
\end{pmatrix}.
\end{eqnarray}
For the association parameters, the following real values were considered: 
\begin{enumerate}
  \item[I.1] $\bm{\alpha}=(-0.5, -0.5, -0.5)$
  \item[I.2] $\bm{\alpha}=(0, -0.5, -0.5)$
  \item[I.3] $\bm{\alpha}=(0, -0.5, -1)$
\end{enumerate}
  \item[Scenario D with dependent markers:] 
 The dependent markers were generated with the following   covariance matrix of the random effects:
  $$\bm{B}=\begin{pmatrix}
\bm{B}^* & \bm{B}^\dagger & \bm{B}^\dagger \\
  & \bm{B}^* & \bm{B}^\dagger \\
 &  & \bm{B}^*
\end{pmatrix},$$

where $\bm{B}^*$ was defined in \eqref{c} and
\begin{eqnarray}\label{c11}
\bm{B}^\dagger=0.5\begin{pmatrix}
1 & 1\\
1 & 1  
\end{pmatrix}.
\end{eqnarray}
Also, the real values of the association parameters were  the same as those considered in scenario I, so we called them D.1, D.2, and D.3 instead of I.1, I.2, and I.3, respectively.
\end{description}

\subsubsection{Scenario 2:  Two Gaussian  and one binary markers }
In this scenario, we considered $K=3$ markers such that the first two markers were Gaussian and the last one was binary. The values of $\bm{\beta}_k$,  $\sigma_1^2$ and $\sigma_2^2$ were the same as the previous scenario, and the structure of the covariance matrix and $\bm{\alpha}$ was considered the same as scenario D. We referred to these structures as M.1, M.2, and M.3 (M for mixed markers), corresponding to 
the three different sets of  $\bm{\alpha}$. 

\subsubsection{Scenario 3: Seven markers }
In this scenario, we considered seven markers such that 
in scenario S.1, the   real values for parameters were similar to the previous scenarios, while, in scenario S.2,  a bootstrapping simulation \cite{hossain2022comparison} was used to mimic PBC2 data. 

\begin{description}
  \item[S.1:] The first marker was considered to be binary, and the others were Gaussian. Also,  $\bm{\beta}_k=(\beta_{0k},\beta_{1k})=(0,-1),~k=1,\cdots,7$ and $\sigma_k^2=0.5,~k=1,\cdots,7$, the  structure of the covariance matrix of the random effects  was considered as
  $$\bm{B}=\begin{pmatrix}
\bm{B}^* & \bm{B}^\dagger & \bm{B}^\dagger & \cdots & \bm{B}^\dagger \\
          & \bm{B}^*      & \bm{B}^\dagger & \cdots & \bm{B}^\dagger \\
          &             & \bm{B}^*     & \cdots & \bm{B}^\dagger \\
          &              &            &  \ddots  & \vdots \\
           &             &             & & \bm{B}^* 
\end{pmatrix},$$
where $\bm{B}^*$ 
and $\bm{B}^\dagger$
were defined in equations \eqref{c} and \eqref{c11}, respectively,  and    $\bm{\alpha}=(-0.5, -0.5, -0.5, 0, 0, 0, 0)$. Thus the first three markers only were independently associated with the event.  
\item[S.2:  Mimicking PBC2 data]  
Here, a bootstrap simulation was applied to mimic data generated in the same structure as PBC2 data to ensure that the data generation mechanism did not depend on any special model structure. For this purpose, $100$ bootstrap samples from PBC2 data were generated. 
\end{description}

\subsubsection{Scenario 4: Time-dependent effects of the markers}
Scenario 4 aims at evaluating the model averaging approach when the effects of the markers on the event risk are time-dependent. Thus, the data generation model for scenario 4 included only two independent longitudinal markers, but their effects on the event risk increased or decreased linearly with time according to the following time-to-event sub-model:
$$\lambda(t)=\lambda_0 \exp (\sum_{k=1}^2 (\alpha_{0k}+ \alpha_{1k}t) m_{ik}(t)).$$
with  $\alpha_{01}= -0.8, \alpha_{11}=0.4$ and  $\alpha_{02}= 0.0, \alpha_{12}=-0.4$, so that the effect of marker 1 is $-0.8$ at $t=0$ and reached 0 at the end of the study ($t=2$) and the reverse for marker 2. Parameters of the mixed models are $\beta_{01}=0.13,\beta_{11}=-0.76$  and $\beta_{02}=0.18,\beta_{12}=-0.62$ and  $\sigma_1=0.56, \sigma_2=0.65$ and the block diagonal elements of $\bm{B}=\begin{pmatrix}
{\bm{B}}_1 & \bm{0}\\
 \bm{0} & {\bm{B}}_2  
\end{pmatrix}$ are: 
$$
{\bm{B}}_1=
\begin{pmatrix}
0.69& 0.01\\
0.01& 0.26\\
\end{pmatrix},
\quad
{\bm{B}}_2=
\begin{pmatrix}
0.74& -0.01\\
-0.01&  0.20\\
\end{pmatrix}.
$$
For scenario S.2 and scenario 4, predictive abilities of the various methods were compared according to the Brier Score instead of the MSE since the true probability of event was unknown in scenario S.2 and not computable with \texttt{JMBayes}  in scenario 4.

\subsection{Results of simulation}
Figures \ref{figc}, \ref{figm}- \ref{figs} and \ref{fig4_1} as well as Tables B. 1-B. 6, B. 7-B. 9, B. 10-B. 13 and B. 14 show the values of $\widehat{AUC}(s,t)$ and $\widehat{BS}(s,t)$ or $\widehat{MSE}(s,t)$ for the landmark time and for different methods and scenarios. 

\begin{itemize}
    \item In the one-marker joint models, when all variables are Gaussian, the markers with the highest association parameter have the best predictive abilities, that is, they have the highest AUC and lowest MSE, regardless of $s$. As expected, the well-specified all-marker joint models have the best predictive abilities. Also, the predictions computed by the one-marker MA at all landmark times exhibit AUC and MSE close to the all-marker JM and better than the best one-marker joint model. When the markers are dependent (D.1, D.2, and D.3), the one-marker MA's performances are slightly better than its performances for independent markers (I.1, I.2, and I.3), since the values of AUC and MSE for the model averaging are closer to those of the all-marker JM. In this scenario, the two-marker MA's individual predictive ability performances are frequently superior to one-marker joint models, two-marker joint models, and the one-marker MA, and comparable to all-marker JM.
    \item One of the markers in scenario 2 is binary, which is less informative than the continuous markers. As a result, the AUC and MSE for the
    one-marker joint model including this binary marker are not as good as those for the continuous one-marker joint model with an equal $\alpha_k$ value. However, in terms of overall performance, the results of scenario 2 show that the model averaging methods, especially the two-marker MA, as a whole perform well. 
    \item For the first scenario with seven markers (S.1, Figure 5), the AUCs of the model averaging methods are very close to those of the all-marker JM and their MSEs are at least as good as those of the best model with the same number of markers. The two-marker model averaging has better performances than its one-marker counterpart. In scenario S.2 mimicking PBC2 data (Figure \ref{figs}), we observe that the performances of the one-marker model averaging are slightly better that those of the all-marker JM. These results may be the consequence of the lack of fit of the all-marker joint model that assumes constant dependent structure between the event and the markers while the MA relaxes this assumption thanks to the time-dependent weights. This situation is investigated in the last simulation scenario.
\item In scenario 4 (Figure \ref{fig4_1} and Table B. 14), the two-marker joint model is misspecified since the association parameters between the event risk and the markers are assumed constant, whereas data are generated with time-dependent associations. Consequently, the two-marker model has better predictive performances than the one-marker models only for $s=0.5$ and $1.0$. In contrast, the model including only marker 1 is better for $s=0$, and the model including only marker 2 is better for $s=1.5$. As the weights in the model averaging method are time-dependent, this method can handle the time-dependent effects of the markers, and it exhibits AUC and BS similar to the best models, whatever $s$.

    \item For a comparison between the computational times for the model averaging methods and the all-marker JM, as we can run the model averaging in parallel, the use of the model averaging methods, especially the one-marker MA, reduces the computational times.

      \item {To assess the effectiveness of the Buckland et al.  \cite{buckland1997model}'s approach in calculating the $95\%$ confidence interval of risk prediction of the MA, we examined the simulated data from scenario 2. The mean of the coverage rates and the mean of the length of intervals for the one-marker MA, two-marker MA and  the credible interval for all-marker JM, estimated by \texttt{JMbayes}, are presented in Table B. 15.  
      The results show that the coverage rates for the MA are acceptable, and in most cases, the coverage rates of the two-marker MA are better than those of the one-marker MA. Also, the coverage rates of the all-marker JM are not as good as those obtained for the MA. }
\end{itemize}

\FloatBarrier
\begin{landscape}
\begin{figure}[ht]
\centering
\includegraphics[width=24cm]{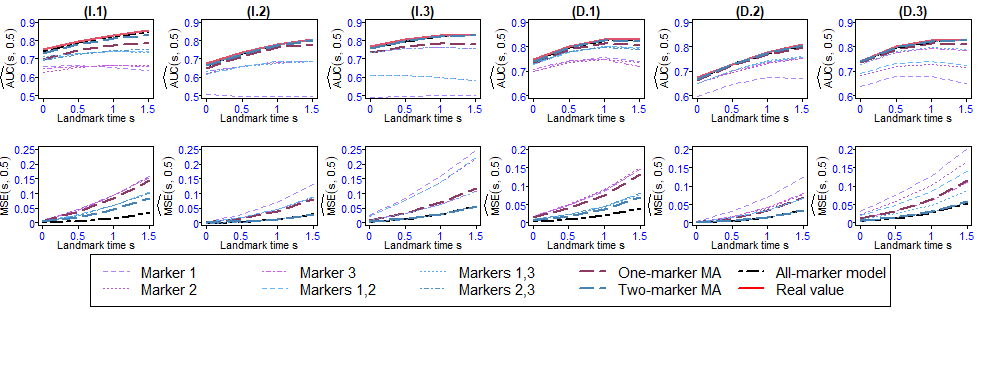}
\vspace*{-1.7cm} \caption{\label{figc}  
$\widehat{AUC}(s,t)$ (first row)   and 
$\widehat{MSE}(s,t)$ (second row) for Scenario 1 with landmark times $s=0, 0.5, 1, 1.5$ and prediction windows of $t=0.5$ over 100 replications. I.1: $\bm{\alpha}=(-0.5, -0.5, -0.5)$ and independent markers, I.2: $\bm{\alpha}=(0, -0.5, -0.5)$ and independent markers, I.3: $\bm{\alpha}=(0, -0.5, -1)$ and independent markers, D.1: $\bm{\alpha}=(-0.5, -0.5, -0.5)$ and dependent markers, D.2: $\bm{\alpha}=(0, -0.5, -0.5)$ and dependent markers and D.3: $\bm{\alpha}=(0, -0.5, -1)$ and dependent markers. }
\end{figure}
\end{landscape}

\FloatBarrier
\begin{figure}[ht]
\centering
\includegraphics[width=16cm]{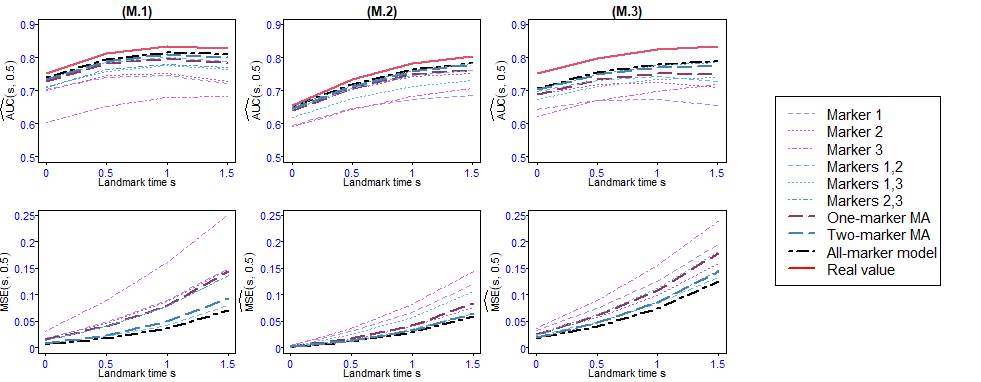}
\vspace*{-.1cm} \caption{\label{figm}  
$\widehat{AUC}(s,t)$ (first row)  and 
$\widehat{MSE}(s,t)$ (second row) for Scenario 2 with landmark times $s=0, 0.5, 1, 1.5$ and prediction windows of $t=0.5$ over 100 replications. M.1: $\bm{\alpha}=(-0.5, -0.5, -0.5)$ and dependent markers, M.2: $\bm{\alpha}=(0, -0.5, -0.5)$ and dependent markers and M.3: $\bm{\alpha}=(0, -0.5, -1)$ and dependent markers.}
\end{figure}

\FloatBarrier
\begin{figure}[ht]
\centering
\includegraphics[width=17cm]{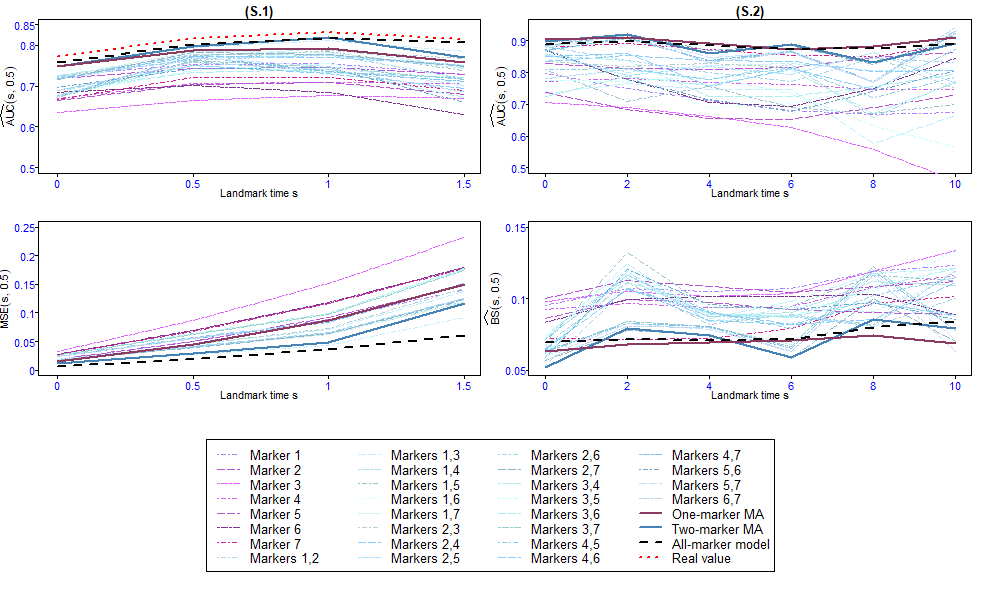}
\vspace*{-.1cm} \caption{\label{figs}  
$\widehat{AUC}(s,t)$ (first row), $\widehat{BS}(s,t)$   or 
$\widehat{MSE}(s,t)$ (second row) for Scenario 3  over 100 replications. S.1: $\bm{\alpha}=(-0.5, -0.5, -0.5, 0, 0, 0, 0)$ and S.2: $\bm{\alpha}=(-0.5, -0.5, -1, 0, 0, 0, 0)$ and dependent markers with landmark times $s=0, 0.5, 1, 1.5$ and prediction windows of $t=0.5$.  S.3: mimicking the PBC2 data by using  bootstrap simulation with landmark times $s=0, 2, 4, 6, 8, 10$ and prediction windows of $t=2$ over 100 replications.
 }
\end{figure}

\FloatBarrier
\begin{figure}[ht]
\centering
\includegraphics[width=14cm]{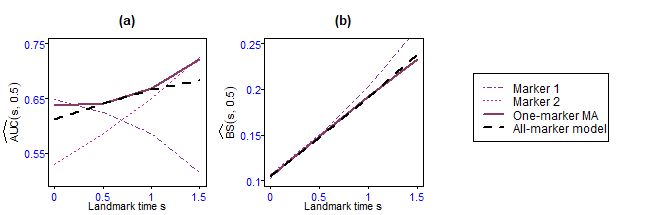}
\vspace*{-.1cm} \caption{\label{fig4_1}  
$\widehat{AUC}(s,t)$ (panel a) and $\widehat{BS}(s,t)$ (panel b)  for Scenario 4  with landmark times $s=0, 0.5, 1, 1.5$ and prediction windows of $t=0.5$ over 100 replications. }
\end{figure}

\section{Discussion}
In this work, we propose a model averaging approach to combine dynamic predictions from several joint models for one event and one or two longitudinal markers. This makes it possible to predict the event risk using the information brought by repeated measures of many markers without estimating a joint model with multiple markers. Indeed, although software programs are available, the estimation of such models is often intractable when the number of markers  increases, particularly with models that include flexible modeling of individual trajectories or non-Gaussian markers.

We estimated the weights for averaging the model-specific prediction probabilities by minimizing the Brier score of the averaged probabilities on the learning data set for each chosen time of prediction and each horizon of prediction. This method has two assets. It accounts for the non-independence between the markers and for the possible time-dependence of the association between the event risk and the marker. When the all-marker joint model is well-specified and may be fitted without numerical issues, simulations showed that model averaging has predictive abilities close to the all-marker joint model. This means that the proposed method can be a good solution when the all-marker joint model is intractable. Moreover, as the weights are time-dependent, this model averaging approach outperforms the all-marker joint model when the dependence structure between the event and the markers is time-dependent. This is a great advantage given that most packages for estimating joint models are not flexible enough to estimate joint models with time-dependent dependence structures. A time-varying association structure was proposed,  \cite{andrinopoulou2018improved}, but this added numerical complexity to the estimation process. The proposed model averaging method can be used regardless of the type of joint models (shared random effects and latent classes) and the computing method for the predictions (classical or Bayesian). 
{ In this paper, we have thoroughly examined the efficacy of the Brier score for deriving weights in dynamic prediction. As a promising avenue for future research, we suggest considering the expected prognostic observed cross-entropy (EPOCE) proposed by Commenges et al.   \cite{commenges2012choice} as an alternative metric. This could offer a valuable addition to the field, potentially enhancing the accuracy and robustness of dynamic prediction models.}

{ In the two real data applications, the predictions obtained by MA are better than those obtained by landmarking with LOCF. The landmark approach is the quickest and simplest approach for dynamic prediction and shares with our MA method the flexibility of time varying association structure, since a time-to-event model is estimated for each time of prediction. However, its performances are deeply impacted by measurement error and length of time between measurements. On the 3C data analysis, MFPCCox exhibited predictive abilities close to the MA approach while FPCA is not expected to be robust to informative dropouts. However, MFPCCox has other drawbacks for some clinical applications, such as its limitation to continuous markers only.
As the FPCs are combinations of
 markers, this method does not allow to identify the most predictive markers. Finally, the risk of event depends on time-fixed summary of the marker trajectories (the individual scores) while, in medical applications, it is most often expected that the risk evolves with change over time of the markers.
On the other hand, MFPCCox is much less computationally demanding than MA.As our objective was to propose an alternative to the full multi-markers joint models, we only compared with landmark and two-step approaches using a proportional hazard model for predicting the time-to-event. Note however that methods combining random survival forests with mixed models or FPCA have also been proposed for dynamic prediction  \cite{devaux2022random,lin2021functional}. RSF may be useful when non linear effects or interactions between markers are expected but comparison with such approaches is outside the scope of this paper. }

Among the two MA approaches we proposed, MA of two-marker joint models often has better performances than its one-marker counterpart because it better accounts for the dependence between the markers but it is at the price of much more computation time when the number of markers is large. On the other side, the one-marker MA has better performances when the dependence structure between the event and the markers may be misspecified. However, the time-dependence of the weights may limit the use of the prediction tools developed with this approach since they could be applied only for the times of prediction considered when computing the weights in the development phase on the learning data set. This limitation is common with the landmark approach. In clinical application, this will require a careful choice of the time-scale (age, time since a clinical event, etc.) and computation of the weights for a large set of prediction times. This problem could be solved by computing weights that are fixed in time or over time intervals. They could be estimated by minimizing the sum of the Brier score contributions for all prediction times or over time intervals. The piecewise constant option is probably a good compromise because allowing time-dependence is a very useful flexibility. Rizopoulos et al. \cite{rizopoulos2014combining} proposed subject-specific weights for averaging predictions from joint models with different dependence structures between the event and the marker. This raises major practical drawbacks since such a prediction tool cannot be applied to new subjects and cannot be validated on an independent data set.

{To make the estimation of the weights more robust, we tried to estimate the weights by minimizing the Brier Score computed within a K-fold cross-validation procedure. However, this increased considerably the computation time (since each one-marker joint model was estimated K fold) without improving significantly the predictive abilities on the validation set in a preliminary simulation results (not shown).}

To conclude, the proposed model averaging approach could contribute to the development of precision medicine by allowing the combination of information from repeated measures of multiple longitudinal markers to compute individual predictions of clinical events. Other weighting approaches could be considered in future work, and their predictive abilities could be compared.

\section*{Acknowledgements}
This work was partly funded by the French National Research Agency (grant ANR-21-CE36 for the project “Joint Models for Epidemiology and Clinical research”). This study was carried out in the framework of the University of Bordeaux's France 2030 program/RRI PHDS.
We thank Christophe Tzourio and Catherine Helmer for providing us with access to the data from the 3C Study.
\\
The 3C Study was supported by Sanofi-Synthélabo, the FRM, the CNAM-TS,DGS,Conseils Régionaux of Aquitaine, Languedoc-Roussillon, and Bourgogne; Foundation of France; Ministry of Research- INSERM ``Cohorts and biological data collections" program; MGEN; Longevity Institute; General Council of the Côte d’Or; ANR PNRA 2006 (grant ANR/ DEDD/ PNRA/ PROJ/ 200206–01-01) and Longvie 2007 (grant LVIE-003-01); Alzheimer Plan Foundation (FCS grant 2009-2012); and Roche.
The Three City Study data are available upon request at e3c.coordinatingcenter@gmail.com.

{\footnotesize{
\bibliographystyle{plain}
\bibliography{wileyNJD-AMA.bib}}}

\newpage
\begin{table}[ht] 
 \caption{\label{auc3c} AUC and Brier score computed by validation set  for landmark times $s=0,5,10$ years and prediction window of 5 years for the 3C data.  }
\centering
\begin{tabular}{c|c|c|c}
  \hline
  &  & AUC & BS   \\ \hline
 & s  & Est. (Sd.) & Est. (Sd.)  \\ 
   \hline
One-marker MA & $0$ &  0.678 (0.055)  &  0.037    (0.007) 
\\ 
              & $5$ &  0.734 (0.038) &  0.073     (0.008)    \\ 
              & $10$ & 0.742 (0.051) &  0.192     (0.017) \\   \hline
MFPCCox      & $0$  & 0.667 (0.054)   & 0.038 (0.007)    \\ 
             & $5$  &  0.668 (0.042)  & 0.076      (0.008)  \\ 
             & $10$ & 0.732 (0.049)   & 0.186     (0.017)  \\  \hline
Landmarking  &   $0$ &  0.628 (0.049) & 0.048 (0.009)  \\ 
  (LOCF)          & $5$ &  0.619 (0.043) & 0.083     (0.096)  \\ 
            & $10$ & 0.714 (0.047)  & 0.193 (0.022)  \\  \hline
pencal  &   $0$ &   -  & - \\ 
            & $5$ &  0.726 (0.039)  & 0.073    (0.008) \\ 
            & $10$ &  0.651 (0.052)   & 0.209 (0.021)  \\  \hline
DynForest  &   $0$ & 0.500 (0.062)  & 0.078 (0.005)   \\ 
            & $5$ & 0.639 (0.043)  & 0.111  (0.006)  \\ 
            & $10$ &  0.621 (0.053) &  0.195 (0.014) \\  \hline  
\end{tabular}
\end{table}
\end{document}